\documentclass{ar-1col} %
\usepackage{natbib} %

\usepackage{enumerate}
\usepackage{amsthm}
\usepackage{amssymb}                 
\usepackage{amsfonts}
\usepackage{array,color}
\usepackage{xcolor}

\usepackage{mathrsfs}
\usepackage{graphicx}
\usepackage{float}
\usepackage{subfigure}
\usepackage{diagbox}
\usepackage{mathtools}
\usepackage{cases}
\usepackage{multicol}

\usepackage{times}

\newcommand{\bx}{{\mathbf x}}

\newcommand{\beq}{\begin{equation}}
\newcommand{\eeq}{\end{equation}}
\newcommand{\beqn}{\begin{eqnarray}}
\newcommand{\eeqn}{\end{eqnarray}}

\newcommand{\pr}{\mbox{Pr}}

\newcommand{\yobs}{Y_{\text{obs}}}

\newcommand{\ymis}{Y_{\text{mis}}}

\numberwithin{equation}{section}

\begin{document}
\markboth{Craiu, Gong and Meng}{Statistical Senses}

\title{Six Statistical Senses}
\author{Radu V. Craiu,$^1$ Ruobin Gong,$^2$ and Xiao-Li Meng$^3$
\affil{$^1$Department of Statistical Sciences, University of Toronto, Toronto, Canada, M5G 1Z5; email: radu.craiu@utoronto.ca}
\affil{$^2$Department of Statistics, Rutgers University, Piscataway, NJ, 08854; email: ruobin.gong@rutgers.edu}
\affil{$^3$Department of Statistics, Harvard University, Cambridge, MA, 02138; email: meng@stat.harvard.edu}}

\begin{abstract}
This article proposes a set of categories, each one representing a particular distillation of important statistical ideas. Each category is labeled a ``sense" because  we think of these as essential in helping every statistical mind connect in constructive and insightful ways with statistical theory, methodologies, and computation, toward the  ultimate goal of building statistical phronesis.  The illustration of each sense with statistical principles and methods provides a sensical tour of the conceptual landscape of statistics, as a leading discipline in the data science ecosystem. 

\end{abstract}


\maketitle

\tableofcontents

\section{What \textit{Are} Statistical Senses?}\label{sec:senses}

\subsection{Statistical Sentience}

Born at the intersection of the empirical  and mathematical universes, statistics needed  time to develop its own set of principles  and tools, intuitions, blunders and  near misses, or in other
terms, its ``character''.  This article  selectively reviews some of the unique features 
that give statistics its disciplinary identity and strength,  and statisticians a toolbox they can lean on in times of creative need.

When writing about important ideas in almost any discipline, one can choose  different paths. There is the genesis trail  which  considers the emergence of statistics as a field and the ideas that propelled it forward \citep{hacking2006emergence,agresti2013strength,shafer2019pascal,shafer2022so,agresti2021foundations}. There is also the historical perspective, excellently illustrated  by \cite{stigler1986history, stigler2002statistics}, in which  chronology  plays an important role in  explaining the evolution of a field. Reading through the history, one cannot avoid the thought that statistics' existence largely preceded the revelation of its essence. The latter concept itself is vulnerable to subjectivity and, had this article been written by a different group, it would have likely resulted in a different creature with different emphases and supported by different interpretations. 

To us, the overarching principle of statistics is linked with the idea of a probability distribution, known, hypothesized or latent. Inexorably linked with a distribution is the concept of variation which, once any effort towards understanding or measuring it is undertaken, leads us to data. Variations permeate our lives because they create simultaneously information and uncertainty; one we love and the other we hate, with our sentiments being also affected by context \citep{Meng2020Information}. Consequently, variations occupy the attention of the discipline of statistics -- and much of the broader data science ecosystem -- by perpetually challenging statisticians and data scientists' ability to separate signals from noise, or even define them. More broadly, for humans to successfully navigate the world of variations and its associated hazards, we have developed our senses to handle the physical reality but also to learn from similar occurrences so as to anticipate the future (prediction) or find rational explanations of the real world (inference). 

For instance, we humans understand that anything can happen tomorrow, but some events are more likely than others. This knowledge is derived from many ``tomorrows'' throughout human history.
Tomorrow  is always different from today,  but it never fails to resemble today to some degree.  Such different-yet-similar repetitions continually shape and enhance our cognitive abilities to recognize patterns, contemplate consequences, cope with uncertainties, and ultimately keep ourselves alive as a species and as individuals. The statistical calculus is fueled by this  fluid tension between similarity and difference. Both depend on one another to possess meaning, and both require nontrivial formulations in a rigorous sense. 
A \textit{probability distribution} prescribes  similarities among individuals in a population by describing their differences. An overarching aim for statistical learning is to pin down, from vastly incomplete information, the most suitable set of distributions that can adequately capture the similarity and difference expressed in available datasets for addressing particular substantive inquiries, confirmatory or exploratory, or a mixture of the two.   
\begin{marginnote}
\entry{Distribution}{a mathematical bookkeeping of individual states by their relative abundance or salience (or the lack thereof).}
\end{marginnote}

In their attempt to systematically advance this learning ability, statisticians have developed over time an arsenal of tools that led to a kind of statistical sentience that inherently defies delineation.
Yet its formation and evolution is undeniable and relies on a number of principles and techniques that, altogether, have proven elemental in developing the tool we need to navigate the world of variations. The task of a statistical learning can seem overwhelming at first.
Based on a sample whose size  is often tiny compared to the population at large, one is asked  to identify complex mechanisms and/or predict future outcomes. Being able to address this enormous problem in all its manifestations would be equivalent to getting a cosmic ``free lunch''. In the absence of the latter, we have to cook ourselves a number of affordable ones to sustain us in different situations.

This aim is served best by developing fundamental  statistical ideas which are time honored and timely, with their values to human inquiries particularly important to emphasize and to realize. This article is devoted to them.
We focus on \emph{statistical ideas} that are the building blocks and milestones for establishing and sustaining statistics as a scientific discipline.  
We connect and highlight these ideas by revealing   %
their roles in forming \emph{statistical senses}, with each one contributing towards  \emph{statistical sentience}. A precise definition for each sense is neither possible nor desirable,
since forcing such a framing would necessarily reduce their rich complexity or inter-connectivity. Nor should we partition a statistician's arsenal into distinct categories, 
because a given technique or principle can appear under different guises or present different merits, and each of them could be serving or enhancing a different sense. Instead, we choose a few distinct concepts and ideas to illustrate each sense.
\begin{marginnote}
\entry{Sentience}{the ability to feel or perceive things.}
\end{marginnote}

This gives us the opportunity to revisit some  cornerstones of our discipline and, more interestingly, to form new links between classical and modern statistics, between ideas and principles that might look disparate at first. By doing so, we hope to facilitate those who are interested in enhancing their statistical sentience professionally or personally,  an ability which may come handy in dealing 
and coping with uncertainties and risks in our increasingly multi-faceted and volatile societies, with the COVID-19 pandemic serving as a painful and prolonged reminder.

\subsection{Statistical Senses}
{In a nutshell, statistical senses are nothing but common senses developed and guided by statistical insights and probabilistic reasoning. However, some seemingly obvious intuitions, such as the expectation that more data must lead to better results, turn out to be statistical fallacies. This is because data contain both signal and noise. Without properly parsing the two, more data may well inflate the noise instead of enhancing the signals. The development of an individual's statistical senses, therefore, is a maturing process that requires time and experiences. The approximate alignment of  the statistician's senses with the natural five senses is intentional, and it is meant to  reify this process.} Just as our natural senses help us navigate and master the physical world, so do the statistical senses help in the world of variations. Indeed, the statistical discipline, as an intellectual body, evolves and thrives because of external stimuli.

The first sense has to do with the {\it informative ignorance} statisticians embrace, which leads us to characterize it as   ``selective hearing''. The latter allows us to get to the heart of a problem without getting lost in details that are less relevant and might greatly complicate the job. For instance, when setting up a regression model, we must decide which features should be included as predictors or explanatory variables, and which ones should be considered as noise. %
Using automated machine learning algorithms to choose features does not release us from making informed judgement.  To the contrary, it only increases the demand on our hearing sensitivity and sensibility, in order to prevent us from being deafened by the machine noise.   

The second sense codes our ability to  ``smell" when a problem is too complicated to solve without first making it seemingly even more complicated, that is, by introducing randomness. Matching two groups of patients on all factors that can affect treatment efficacy is an impossible task. Yet by
flipping a fair coin to randomly assign each patient to a treatment or a placebo, we ensure the balance \textit{statistically} between the two groups on all known, unknown, or unknowable factors, from which a causal inference of the treatment efficacy becomes possible. Fully protecting data privacy is another impossible task. Yet by injecting properly designed randomness into data before releasing, we can control the level of privacy loss while maintaining reasonable utility of the data.

The third sense is a form of enhanced vision, where experienced statisticians are able to see through data that are not present or, more broadly, \textit{dark data}. Competent detectives do not reason only from what items are present at a crime scene. A missing portrait can be a smoking gun more than a gun smoking. Most times data do not miss randomly. Many individuals surveyed during the 2016 US presidential election did not decide to not respond by flipping a fair coin. We all have seen consequences of such survey data being analyzed and reported without the benefit of a ``seeing through" sense.

The fourth sense, which we term  the ``magic touch", concerns  statistical principles and insights that can convert  ordinary estimators into  extraordinary ones, literally and figuratively.  A shining example is the so-called "Rao-Blackwellization", which turns a noisy estimator into an optimal one by a form of deep stratification, or more broadly, \textit{conditioning}. Conditioning, in  layperson's terms, means to take into account more detailed information that are judged to be relevant, just as a patient wants an effective individualized treatment, instead of one
that works on average. The magic touch of conditioning requires both experience and careful thinking. The famous Monte Hall problem,  regarding whether one should switch a door or not in order to win a prize behind the door, demonstrates how human minds can easily be misled by a lack of training on probabilistic conditioning.    

Extracting  meaningful information from data is perhaps the most important part of a statistician's \emph{raison d'\^etre}, so it is not surprising that the fifth sense is defined by an ability to extract a wealth of information from the frugality of a sample (relative to the much larger population), or ``just a taste", the fifth statistical sense. In Occam's razor, statisticians discovered a principle that has guided and served them for many years. The richness that comes from frugality shines through other illustrious instances that range from  bootstrap to propensity matching. But there is no free lunch. Bootstrap is made possible by the hidden assumption that there are inherent replications within a single data set, and propensity matching works well  when there is sufficient built-in compatibility between the samples being matched. A key part of statistical sentience is to have a good sense of the limitations of each method, now matter how almighty it may appear.

This leads to the sixth (statistical) sense, which is meant to express exactly that: an ability to transcend pure rationality or technical prowess and to create a qualitative jump that is as fallible as it is path-breaking. As we shall argue and illustrate, all attempts to transition from known to unknown require a leap of faith, explicit or concealed, since statistical contemplation by its very nature involves building shaky bridges that come with structural risks. Much statistical training is needed to build a ``sixth sense'', after acquiring which one would be daring enough to spearhead new methods despite the inherent risks.  Upon completing our sensical tour,  one sense per section, we explain in a final section the origin of this tour, how our senses have evolved, and invite our readers to help  promote sensical statistics and data science as we venture deeper into the digital age.

\section{Selective Hearing: Informative Ignorance}\label{sec:info}

Statisticians must be able to hear the essential and ignore the incidental in their data so that they can develop statistical models with the relevant inferential aims. The latter could be the essential effects of a treatment in randomized studies or disentangling the persistent similarities from spurious differences using de Finetti's elegant exchangeability result.

\subsection{Randomized Replications}

To provide a concrete example of how randomized replications and probability distributions work in practice, consider the problem of comparing the resilience of $K$ different brands of tires. A total of $M$ drivers are selected at random and are asked to test-drive all brands for a number of weeks.  Let  $Y_{ij}$ denote the wear recorded for brand $i$ by driver $j$. The $Y_{ij}$'s  differ from each other, but intuitively those sharing the same driver or  the same brand should be more similar than those which share neither. %
We may capture such similarities and differences by positing the following distributional model for $Y_{ij}$: 
\beq
Y_{ij} = a+ b_i + d_j + \epsilon_{ij},
\label{lm}
\eeq
 where $a$ represents a common baseline,  and $b_i$ and $d_j$ aim to capture the effects of using brand $i$ and driver $j$, respectively. The so-called error term $\epsilon_{ij}$ describes the idiosyncratic variabilities, which are often assumed to form an independently and identically distributed (i.i.d.) sample from  $N(0,\sigma^2)$, with $\sigma^2$ to be estimated from the data. 
 \begin{marginnote}
 \entry{i.i.d}{A set of random variables that are mutually independent of each other, but share the same probability distribution. }
 \end{marginnote}
 
This model alone does not capture similarities among drivers or brands. The fact that they are all persons or tires implies that they have similarities.  We can again use distributions to model their similarities, such as 
 \begin{align}
b_1,\ldots,b_K\stackrel{i.i.d.}{\sim} N(0, \tau^2_b),  \quad 
d_1,\ldots,d_M \stackrel{i.i.d.}{\sim} N(0, \tau^2_d),
 \label{lmm}
 \end{align}
where $\{b_1, \ldots, b_K\}$ and $\{d_1, \ldots, d_M\}$ are independent of each other, and the variances $\tau_b^2$ and $\tau_d^2$ are to be estimated from the data.  The specification given by \eqref{lm}-\eqref{lmm} is a special case of the so-called random-effects model, whose essence is to describe a randomized replication study via a probabilistic distribution. Fitting and checking such models constitutes a large portion of statistical endeavors, yet setting up the appropriate randomized replications is always the most critical step, be it a thought experiment or a real one. Without a (thought) mechanism that permits us to ignore in order to inform, we would literally drown in the data tsunami, or even a data stream.

If instead of wear, we record the time, $T_{ij}$, of the first failure of a brand $j$ tire driven by driver $i$, then studying the dependence between the response and  say, a  covariate vector $\bx_{ij}$  using model \eqref{lm} is inadequate. Estimates of regression coefficients are  difficult to interpret and could be biased because the model ignores important features of the data structure, e.g., censoring. For instance, the latter occurs if some of the drivers do not experience a tire failure before the study is completed. The proportional hazards (PH) model of \cite{cox1972regression} (see sidebar), 
developed precisely for studying the covariate effects on a time-to-event response variable, is another example of a sentient statistician's selective hearing. Cox's model assumes that the \emph{ratio} of hazards  associated to two different brand/driver pairs  depends solely on the covariate vector and is constant in time.  An added benefit is that the modeler is freed from the {essentially impossible  task of correctly specifying  the entire baseline hazard function. }
This ingenious construction has turned the PH  into one of the most widely used statistical models due to its versatility which allows, among other things, time-dependent covariates \citep{fisher1999time} and   the integration of  random effects \citep{vaida2000proportional}.

\begin{textbox}[t]\section{Cox's Proportional Hazards Model}
The proportional hazards (PH) model  \citep{cox1972regression} allows the estimation of covariate effects on   time-to-event response variables. It has found many applications in biomedical studies of survival and engineering studies of reliability.  The hazard function at time $t$ for a survival time variable $T$, $\lambda(t)$, 
can be interpreted as 
proportional to the instantaneous probability of the event  (death, failure, etc) at  time $t$. In the presence of a covariate vector $\bx$ that is expected to influence the distribution of $T$, the PH model posits that the hazard is the product between  a baseline hazard $\lambda_0(t)$, independent of $\bx$, and a positive function of the linear predictor $\beta^T \bx$ (usually $\exp (\beta^T\bx)$, but other forms are possible).   %

\end{textbox}

\subsection{Exchangeablility}

Statistics offers  expectations and predictions about occurrences that are not observed, based on some that are. What makes an inferential claim \emph{statistical} is that, typically, the observed evidence on which the inference is based consists of a collection of variables which are amenable to a probabilistic description about their similarities and differences. Any reader of statistical textbooks would be familiar with the i.i.d. assumption introduced in the previous section as a foundational element for proving validity of various probabilistic methods.
A collection of independent and identically distributed  random variables exists only in an idealized form, under the assumption that the distribution of outcomes for the repeated events is known to be stable (in time, across subpopulations, etc.), {or when the i.i.d. concept is used as a theoretical tool to prescribe differences and similarities, as in (\eqref{lm})-(\eqref{lmm})}. In reality, the i.i.d. assumption rarely withstands close scrutiny.

Not surprisingly, statisticians and probabilists have been greatly bothered by this compromise. The great insight offered by the exchangeability concept allows statistical analyses to retain many of the  advantages offered by the i.i.d. setup while capturing realistically a much wider range of repeatable random phenomena. Mathematically, the key assumption we invoke to model randomized replications is  that they share the same probabilistic distribution -- or rather, that we have no evidence to claim that they don't. Such failure to distinguish the random quantities from one another beyond their shared probabilistic distribution  grants them the status of being replications to one another. \cite{de2017theory} used the term \emph{exchangeability} to replace what he thought as a misnomer description of ``independent events of unknown probability.''  
A finite number of random variables are called exchangeable if their joint distribution is invariant to permutations. By definition, an infinite collection is exchangeable as long as so are any of its finite subsets. 

\begin{marginnote}
\entry{Exchangeability}{A set of random variables are exchangeable if their joint probability stays the same  regardless how we permute  their labels.}

\end{marginnote}

De Finetti showed that an infinite sequence of  Bernoulli random variables $\{X_n; \; n\ge 1\}$ is exchangeable if and only if there exists a distribution  $F$ such that 
$\{X_n; \; n \ge 1\}$ are  conditionally independent and 
$P(X_1=x_1,\ldots,X_n=x_n)=\int \theta^{S_n} (1-\theta)^{n-S_n}dF(\theta)$, where $S_n=\sum_{k=1}^n x_i$ for all $n\ge 1$. Moreover, $\theta$ is the limiting frequency  
$\theta=\lim_{n \rightarrow \infty}{S_n/n}$.
Prior to collecting data, one could model \textit{a priori} the distribution $F$ using the available knowledge  about the long range behavior of the Bernoulli trials and the conditioning variable $\theta$ emerges naturally as the unknown probability of success. Therefore, the de Finetti theorem provides coherence and theoretical conceptualization  for all the ingredients required in a Bayesian analysis. 
Exchangeability also builds connections with Fisher's concept of subpopulations as noticed in \cite{lindley1981role}, and allows a principled treatment of underlying heterogeneous  structures in a population.

In practice, finding the  right variable to condition on provides clear benefits because  it allows us to think of the data as i.i.d. which, in turn, makes it possible to predict the unseen from the seen. Using  temporal data for illustration,  if one  believes that the future is exchangeable with the past, then the knowledge accumulated from measuring and understanding the past can inform, in probabilistic ways,  the future.  However, this needs careful qualifications,  since tomorrow is clearly not exchangeable with today in a strict sense, minimally because the time is not reversible. What we typically mean is that those aspects of the future which we cannot (reasonably) describe using known characteristics will present uncertainties that are exchangeable with those recorded in the past.

\section{Following Our Nose: Determinacy through Randomness}\label{sec:nose}

 Insertion of randomness in a system is a general principle which, when mastered properly, can overcome complexity in computation or in inference. {There are many rewarding uses of this seemingly counter-intuitive sense, with the aforementioned randomization being an obvious one.   Here we further illustrate its versatility via Monte Carlo and statistical privacy.}

\subsection{Monte Carlo Integration}\label{sec:MC}

Statistical practice often requires investigating the properties of a function $f(\theta)$ when $\theta$ varies in some state space $\Theta$.
Consider the common problem of computing the expectation (i.e., average) of $f$ with respect to a density $\pi(\theta)$: $I=\int_\Theta f(\theta)\pi(\theta)d\theta$. Analytical calculation or deterministic numerical approximation \citep[e.g.,][]{owen2003quasi} often is impractical for many problems, especially
when the dimension of $\theta$ is high. 
The Monte Carlo method \citep{metropolis1949monte} then becomes the only choice, by randomly sampling points $\{\theta_1, \ldots, \theta_n\}$ from $\pi(\theta)$, and then taking average of $\{f(\theta_i), i=1, \ldots, n\}$, denoted by $\hat I_n$, to form the Monte Carlo estimator for $I$. 

Obviously, the quality of the sampled points $\{\theta_1, \ldots, \theta_n\}$ is critical in determining the statistical accuracy of the $\hat I_n$ as an estimator for $I$. An i.i.d. sample from $\pi$  generally is considered ideal but typically is very difficult to obtain even for reasonably looking $\pi$. However, the unbiasedness of $\hat I_n$ is preserved as long as each $\theta_i$ is drawn from $\pi$, regardless how statistically $\{\theta_1, \ldots, \theta_n\}$ may depend on each other. Consequently, we can give up the independence requirement in i.i.d. and construct the so-called Markov chain in the form of $\theta_{t+1}=\psi(\theta_t, U_{t+1})$, where $t$ indexes the iteration,  $\psi$ is a deterministic {\it updating function}, and $U_{t+1}$ is a random vector independent of all its predecessors $\{U_0, U_1, \ldots, U_t\}$. The central idea here is that by carefully choosing the function $\psi$, the resulting chain will generate a sequence of $\theta_t$ whose statistical properties approach those that are generated directly from $\pi$, as $t$ increases. This construction is known as Markov chain Monte Carlo (MCMC), which has revolutionized the Bayesian computation since the 1990s \citep{Geman1984,tanner1987calculation, gelf_smith,gilks1994language,hand-mcmc}.  

MCMC is also an example of a \textit{statistical idea} that was not an idea in statistics  -- that is, neither invented by statisticians nor first published in the statistical literature --  because it was developed in the early 1950s by  %
physicists \citep{Metropolis1953}. Later, building on an analogy between images and lattice physical systems, \cite{Geman1984} introduced the Gibbs sampler (see sidebar) which was soon thereafter used in Bayesian methods for image reconstructions \citep{besag1986statistical,besag1993spatial}, and Bayesian analysis for general statistical models \citep{gelf_smith}, including those formulated by augmenting the observed data with  missing or latent ones \citep{tanner1987calculation}.

\begin{textbox}[t]\section{The Gibbs Sampler}
The Gibbs sampler is an MCMC algorithm which is used to sample from a $d$ dimensional posterior distribution $\pi(\theta)$ %
\citep{Geman1984,liu1995covariance}. After initializing the chain at $\theta^{(0)}$, the chain updates at any iteration $t \ge 1$ by cycling through all the components of $\theta^{(t)}$ and sampling a new value for $\theta^{(t)}_j$ from $\pi(\theta^{(t)}_j\mid
\mathbf{\theta}^{(t)}_{1:(j-1)},
\mathbf{\theta}^{(t-1)}_{(j+1):d})$,
the conditional distribution of the $j$-th component given the values of the remaining components at the current iteration. An important variation is the {\it block Gibbs sampler} which cycle and update groups of components instead of individual ones.

\end{textbox}

The well-known algorithm by \cite{Metropolis1953}, later generalized by \cite{hastings1970monte}, demonstrates  how to construct $\psi$ to sample from a target $\pi$.
We start with an initial $\theta_0$, and then sample $\theta^*$ (independently) from a proposal density $p$, typically chosen as a decent approximation of $\pi$ but much easier to sample from. We then compute $r=\pi(\theta^*)/\pi(\theta_0)$. If $r\ge 1$, then we know that under $\pi$ the value $\theta^*$ should appear more often in our sample than  $\theta_0$. Since $\theta_0$ is already in our sample, we will need to include $\theta^*$ as well,
and hence we let $\theta_1=\theta^*$. If $r<1$, say $r=0.25$, then the value $\theta_0$ should appear in our sample four times more frequently than $\theta^*$. We can achieve  this by flipping a biased coin which lands head $25\%$ of the time, and let $\theta_1=\theta^*$ if it is head, and $\theta_1=\theta_0$ if it is tail. We then repeat this process,  with $\theta_1$ replacing $\theta_0$, to determine $\theta_2$, and subsequently $\theta_t, t=3, 4, \ldots $.

Theoretical guarantees require a set of regularity conditions, under which the distribution of $\theta_t$ from the Metropolis-Hastings, Gibbs, or many other MCMC algorithms, would approach
 the target $\pi$ \citep{Meyn1994}. This means that when $t$ is sufficiently large, say, larger than an integer $B$ which is referred to as the burn-in and can be estimated in various ways \citep{MR1665662, Rosenthal2002}, 
 we can treat $\{\theta_t, t\ge B\}$ as an approximate sample from $\pi$ and hence use them to form the Monte Carlo estimate $I_n$. 

The magic of statistical estimation is not only that it provides an estimator with known theoretical properties, but it also assesses the statistical uncertainty about the estimator itself.    %
A central limit theorem for Markov chains \citep{roberts96, Jones:2001gf} ensures that the distribution of $I_n$ will be well approximated by the normal distribution $N(I,  \hat\sigma^2_f/n)$ when $n$ is sufficiently large. Here, $\hat\sigma^2_f$ is computed from $\{f(\theta_t), t\ge B\}$ \citep{geyer92,flegal,vats2019multivariate} using  
\beq
\hat\sigma_f^2={\hat V_f}\left[ 1 + {2\over n}\sum_{k=1+B}^{n+B}(n-k) \hat\rho_k \right],
\label{finvar}
\eeq
where $\hat V_f$ and $\hat\rho_k$ are the sampling variance and lag-$k$ auto-correlation from $\{f(\theta_t), t\ge B\}$, respectively.    
This would allow us to provide an approximated $95\%$ \textit{confidence interval estimator} for $I$ in the form of $\hat I_n \pm \hat\tau$, where $\hat\tau$ is an estimated margin of error given by $\hat\tau=2 \hat\sigma_f/\sqrt{n}$.

From expression (\eqref{finvar}),  we see Monte Carlo integration avoids the curse of dimension once we know how to sample correctly from the target $\pi$. %
Expression (\eqref{finvar}) also makes it apparent that reducing the within-chain correlations $\rho_k$ will reduce the MCMC variance.   
Interestingly enough, randomness can be again  adapted to serve our goals, for instance by introducing  dependencies in the design of MCMC samplers to yield negative $\rho$'s \citep{Rubi:Samo:85:VRB, FGR, craiu2005multiprocess,craiu2007acceleration}  or even to dissolve the dependence altogether and generate i.i.d. samples, the so-called perfect or exact sampling \citep[e.g., see][]{propp-wilson:exact-sampling,craiu2011perfection}.  The construction of a perfect sampler relies on a clever use of randomness via a coupling strategy of multiple Markov chains,  {which often is not an easy task. Fortunately, similar coupling techniques have led to unbiased estimates of $I$ for general MCMC algorithms \citep{heng2019unbiased,jacob2020unbiased}, which can be viewed as a sensible compromise between being perfect and being practical \citep{craiu2020double}.}

\subsection{Randomized Responses and Differential Privacy}

A challenge to the elicitation of survey response is when the question being surveyed is sensitive in nature. When respondents, especially those with perceived negative attributes (e.g., health condition, smoking, substance abuse), evade responses or lie about them, they induce systematic biases in the resulting statistical inference, and the power of randomization cannot be exercised fully.

The \emph{randomized response} mechanism \citep{warner1965randomized} is an ingenious idea that can alleviate bias due to evasive answers in surveys. It uses a surprising \emph{double} application of randomization within survey sampling. The mechanism, as originally constructed by \citeauthor{warner1965randomized}, is described in the sidebar. Using a random device -- a biased coin with a known probability $p \in (1/2, 1)$ of turning up heads -- to elicit answers, no individual has to report their sensitive feature verbatim, yet these randomized answers allow the interviewer to leverage statistical methods to unbiasedly infer the underlying true proportion and quantify uncertainty.

\begin{textbox}[t]\section{Randomized Response}

 The randomized response strategy of \cite{warner1965randomized} is the earliest known mechanism that satisfies \emph{differential privacy} \citep{dwork2006calibrating}. Let $X_i \in \{0, 1\}$ be a binary attribute for individual $i$, and $\pi$ be the population proportion of 1's, which we wish to estimate.
A total of $n$ individuals are sampled with equal probability and each of them is given a random device $R_i$ (e.g., a biased coin) that simulates a Bernoulli random variable with known probability $p \in (1/2, 1)$. The individual reports $Y_i = 1$ if  $R_i = X_i$, and $Y_i = 0$ otherwise, but does not disclose $R_i$ or $X_i$.
Consequently, the probability of observing $Y=1$ is
$\pi p+\left(1-\pi\right)\left(1-p\right)$. This allows
us to estimate $\pi$ unbiasedly---without knowing any anyone's actual data $X$--- 
via
$\hat{\pi}=\left(p-1+\bar Y_n\right)/\left(2p-1\right)$, where $\bar Y_n$ is the average of the observed $\{Y_1, \ldots, Y_n\}$.
\end{textbox}

The randomized response mechanism is a \emph{differentially private} mechanism \citep{dwork2006calibrating}, a formal privacy concept proposed decades later. Differential privacy has become the state-of-the-art standard for disclosure limitation for major corporations and statistical agencies, as seen in its high-profile adoption by the U.S. Census Bureau for protecting the 2020 Decennial Census, including both the P.L. 94-171 redistricting data released in August 2021 \citep{abowd2022topdown}, and the Demographics and Housing Charateristics data scheduled for release in late 2023  \citep{abowd2023confidentiality}.

To say that a random function $Y$ is $\epsilon$-differentially private means that, {for all neighboring values $(x, x')$  (typically with unit metric in the neighborhood definition) the input data $X$ can take, and for all possible states $y$ for $Y$ (here we assume $Y$ is discrete for simplicity)},
\begin{equation}\label{eq:prob-ratio}
\frac{\Pr\left(Y\left(X\right)=y\mid X=x\right)}{\Pr\left(Y\left(X\right)=y\mid X=x'\right)}\le\exp\left(\epsilon\right)
\end{equation}
for a chosen $\epsilon>0$, known as ``privacy loss budget''. The larger the $\epsilon$, the lesser the privacy protection, with a trade-off of greater preservation of information in the data.  Differential privacy amounts to requiring that the probability distribution of $Y$ does not change too much upon small changes in $X$. We see that the randomized response mechanism is a $\epsilon$-differentially private mechanism with $\epsilon={\rm logit} (p)=\log(p/(1-p))$, because the  conditional probability ratios as (\eqref{eq:prob-ratio}) requires, namely $ {\Pr\left(Y_{i}=1\mid X_{i}=1\right)}/{\Pr\left(Y_{i}=1\mid X_{i}=0\right)}$ and ${\Pr\left(Y_{i}=0\mid X_{i}=0\right)}/{\Pr\left(Y_{i}=0\mid X_{i}=1\right)}$, are both equal to $p/(1-p)$ by design. This also explains why $p=1/2$ will make $\hat\pi$ take the useless value of $\infty$, because consequently $\epsilon=0$, which means that the data are completely private, and hence no information for estimating $\pi$ may come from the data alone. 

On the other hand, when $p$ approaches 1, $\epsilon$ approaches infinity, which means that we will lose all privacy protection. But as a trade-off, we can have the fully efficient estimate $\hat\pi =\bar Y_n$. Of course, this full efficiency is not achievable in practice, because without some degrees of privacy protection, some individuals would refuse to respond. Worse, their decisions to not respond  likely correlate with  answers to sensitive question being asked. This will lead to not only reduced sample size, but most critically to a biased sample (see general discussions of these problems in the next section), the very reason that we want to use a randomized mechanism in the first place. 

The randomized response mechanism, and more general differentially private mechanisms that come after it, are a great example of how a calculated randomness can be used to our advantage, to probe difficult and confidential questions that do not otherwise succumb to direct and deterministic conquering. The transparent specification of these mechanisms allow for the principled decoding of statistical information contained in the veiled responses elicited from individuals \citep{gong0000exact,gong2022transparent}. For an in-depth review of differential privacy and statistical disclosure limitation, see \cite{slavkovic2022statistical}.

\section{Seeing Through: Enlightenment from Dark Data}\label{sec:seeing}

 Data are never completely observed in reality and their hidden parts  have generating mechanisms that can greatly complicate statistical analysis because of the \textit{selection biases} they create.  %
 The statistician's ability to decipher the latter and  conjure ways to complete the former, seemingly out of thin air, has lead to computational shortcuts and to more meaningful analyses.  We illustrate the benefits of this enhanced vision with topics in sample survey and Bayesian computation.
 
 \begin{marginnote}
\entry{Selection Bias}{Systematic distortions created by selection processes in data collection that destroy their representativeness.}
\end{marginnote}

\subsection{Quantifying Missing Data Mechanisms and Data Defects}\label{sec:ddc}

\textit{Dark data} is a general term for any kind of data that are lost or distorted before analysis, as well as for   unobserved or unobservable ``data'' such as latent variables that are constructed or conceptualized because they  serve modeling and computational purposes \citep[e.g.,][]{hand2020dark}. The need to deal with missing or incomplete data, the most common form of dark data,  is a rule rather than an exception. All of us, for example, have ignored survey questionnaires or provided only partial answers multiple times in our lives. 
The missing or incomplete data generally create at least three problems for analysis: (1) deteriorating data quality, (2) reducing data quantity, and (3) impeding the use of standard methods and software \citep{meng2012you}. For example, when the reason for a non-response is correlated  with the answers we are seeking, the survey data we observe are no longer representative of our target population.
The consequence of this data distortion is more devastating than commonly realized.  
\begin{marginnote}
\entry{Dark Data}{Unobserved or unobservable data, but their contemplation may lead to better model or computation.}
\end{marginnote}

To see this clearly, \cite{meng2018statistical} shows that the actual error induced when using a sample mean ($\bar Y_n$) to estimate a population mean ($\bar Y_N)$ can be decomposed into three factors: 
\begin{equation} \label{eq:bib}
 \underbrace{\bar{Y}_n -  \bar{Y}_N}_{\textbf{Actual Error}}
 = \underbrace{\rho}_{{\textbf{Data Quality}}}
 \times 
\underbrace{\sqrt{\frac{1-f}{f}}}_{\textbf{Data Quantity}}
 \times
 \underbrace{\sigma}_{\textbf{Problem Difficulty}}.
\end{equation}
Here $\rho$ is the \textit{finite-population} correlation between $Y_i$ and the binary recording indicator variable $R_i$ (i.e.,  $R_i=1$ if $Y_i$ is recorded in the sample, and $R_i=0$ otherwise), and $\sigma$ is the standard deviation of the finite population $\{Y_1, \ldots, Y_N\}$. The second factor is entirely determined by the relative sample size $f=n/N$, and the third measures the finite population heterogeneity and hence the difficulty of estimating its mean.  The first factor, the \textit{data defect correlation} (ddc) $\rho$,  is a useful measure for data quality because it captures the selection bias created by the dependence of $R_i$ on $Y_i$; for a truly probabilistic sample, $\rho$ is zero on average since whether $R_i=1$ or not is uninfluenced by $Y_i$. 
\begin{marginnote}
\entry{Finite-Population Statistics}{All individual attributes are treated as fixed, and all statistics are formed by enumerating over individual indices in the (finite) population.}
\end{marginnote}

When $\rho$ is not negligible (compared to $N^{-1}$), the resulting mean-squared error of $\bar X_n$ (averaging over possible randomness in $R$) is the same as that of a simple random sample with  size $n_{\rm eff}\approx f/(1-f) \rho^{-2}$. For the 2016 US Presidential election, $\rho\approx -0.005$ pertains to voting for Trump based on  survey data used in \cite{meng2018statistical}. This means that the sample mean from 2.3 million potential voters--about $1\%$ of the US eligible voters--would not be more accurate for estimating Trump's vote share than that from a simple random sampling of about 400 voters who respond fully and truthfully, a 99.98\% reduction from the apparent ``big data''. Hence the importance of data quality cannot be overemphasized.

The seemingly trivial adoption of the recording indicator $R$ reflects a great advance in statistical analysis because it introduces a probabilistic framework
for quantifying and modeling the selection biases created by any kind of missing-data mechanism. Specifically,  \cite{rubin1976inference} used the conditional probability $\Pr(R=1|\yobs, \ymis)$ to model the mechanism, where $\yobs$ denote the observed data points, and $\ymis$ the missing ones. Selection bias exists when $\Pr(R=1|\yobs, \ymis)$ varies with $\ymis$, a situation typically described as suffering from a \textit{non-ignorable} mechanism. When $\Pr(R=1|\yobs, \ymis)\equiv \Pr(R=1|\yobs)$, the so-called \textit{missing at random}, we can avoid selection bias by incorporating $\Pr(R=1|\yobs)$ in the analysis.  The easiest case is when $\Pr(R=1|\yobs, \ymis)$ is not influenced by either $\yobs$ or $\ymis$, in which case $\yobs$ is simply a random sub-sample of the random sample we intended to collect, and hence $\yobs$ can be analyzed using standard methods. {But this is a very rare occurrence in practice, so many more advanced methods have been developed and applied to address the issue of missing data \citep{Rubin1987,little2019statistical,enders2022applied} and more generally the challenges associated wwith a non-probabilistic sample \citep{elliott2017inference,zhang2019valid,wu2022}.}

\begin{marginnote}
\entry{Missing-Data Mechanism}{The process that prevents data from being fully observed, a process often is described by a probability model.} 
\end{marginnote}

\subsection{Intentionally Constructed Dark Data}\label{subsec:potential-outcome}

Rather than regretting their unfortunate occurrence, one may choose to intentionally construct dark data either to aid the expression of the statistical model or to facilitate computation. 
As reviewed in \cite{meng2000missing}, hypothetical data construction come in many forms and shapes, such as latent variables \citep{loehlin2004latent}, auxiliary variables \citep{pollock2002use}, hidden states \citep{elliott2008hidden},  and so on. 

Perhaps a most daring, yet certainly prolific, construction of dark data is the \emph{potential outcome} notation, to refer to the complete set of possible values of a unit under all arms of treatment in an experiment \citep{neyman1923application}. 
Suppose a clinical trial for a vaccine  is designed such that a representative sample from the population is randomly assigned to two treatment arms, with $Z_i = 1$ indicating receipt of the vaccine by individual $i$, and $Z_i = 0$ of the placebo. All individuals complied with the study, and their immunity effects measured and recorded accurately as $Y_i=Y(Z_i)$. 
However, even with such practically unachievable idealization,  there is an inherent missing data structure that prevents us from even defining the treatment efficacy if we do not recognize it. 
In a typical experiment, an individual can receive either the vaccine or its placebo, but not both. But it is exactly the difference of the outcomes from both that defines the efficacy for the individual.  In notation, we are interested in assessing $e_i=Y_i(1) - Y_i(0)$ for individual $i$, which is never directly observable. Nevertheless, the formulation via the potential outcome $\{Y_i(1), Y_i(0)\}$ is critical in clearly defining the scientifically meaningful estimand. It also renders clearer what are estimable and what are not. 

For example, whereas it is impossible to assess individual efficacy $e_i$,  for a finite population of size $N$, we can estimate the  average causal effect,  $A_N= \sum_{i=1}^N e_i/N$ by $\hat A_n=\bar Y(1) - \bar Y(0)$, where $\bar Y(Z)$ is the sample average of the outcomes from all individuals who have received the treatment $Z (=0, 1)$. When the assignment to the treatment is done randomly, which is the case for most clinical trials, $\hat A_n$ is an unbiased estimate of $A_N$, despite the fact that we cannot estimate any individual $e_i$ unbiasedly without making model assumptions (e.g., all individual efficiencies are the same).

Potential outcomes follow naturally from our imagination, and permit the entertainment of \emph{counterfactuals} and \emph{possible worlds} beyond the observable one \citep{lewis1974causation,sep-possible-worlds}. With or without \citeauthor{neyman1923application}'s notation, the idea permeates the study of causal inference from randomized experiments in statistics, economics and the social sciences; e.g., \citet[Book III Chapter 5]{mill1906system}; \cite{fisher1919causes,tinbergen1930bestimmung,haavelmo1943statistical, cochran1957experimental, cox1958planning}. \cite{rubin1974estimating,rubin1978bayesian} established the use of potential outcomes in observational studies, providing a formal framework for the analysis of causal effects that extends beyond classical randomized experiments.

\begin{textbox}[t]\section{Expectation-Maximization (EM) Algorithm}
 
The EM algorithm is used to find the MLE of a parameter, $\theta$, when a portion of the data is missing.  Initialized at $\theta_0$, the algorithm proceeds in an iterative manner over two steps (until convergence under a pre-specified criteria):

\smallskip
\noindent
{\bf E-step}: Compute $\mbox{Q}(\theta \mid \theta^{(t)})$, the conditional expectation of the complete-data log likelihood given the observed data $X$ and $\theta=\theta^{(t)}$, the parameter estimate at the current iteration;

\noindent
{\bf M-step}: Find the $\theta$ value that maximizes $\mbox{Q}(\theta\mid \theta^{(t)})$ and set it to $\theta^{(t+1)}$.
\smallskip

 Emerging first in the works of \cite{baum1970maximization} and \cite{sundberg1976iterative,sundberg1974maximum}, the EM algorithm took shape and became widely popular through \cite{dempster1977maximum}. It remains one of the most influential algorithms of all time.

\end{textbox}

Another kind of intentionally constructed missing data appears in  computational algorithms, known as \emph{data augmentation}, a term coined by \cite{tanner1987calculation} in the context of MCMC, as discussed in Section~\ref{sec:MC}.  One of the most useful illustrations of this principle is represented by the EM algorithm for computing the maximum likelihood estimator (MLE) with missing data.
Suppose that we are interested in computing the MLE for $\theta$ from a log-likelihood $\ell(\theta|X)$,  where $X$ denotes the data we observe. Here $X$ may be only part of an intended larger data set, such as $X=\yobs$ using the notation from before, or it is the entire intended data set.  Nevertheless, we can always augment $X$ to $\{X, Z\}$, since bringing the augmented data $Z$ %
can make finding MLE from the  augmented data likelihood $\ell(\theta|Z,X)$ easier than from our original $\ell(\theta|X)$. On the other hand,  once we have an estimate for $\theta$, we can use the conditional model $f(Z|X, \theta)$ to ``impute/predict'' the $\ell(\theta|Z,X)$ function (since it depends on the unknown $Z$),  which in turn can lead to a better estimate of $\theta$.   The EM algorithm formalizes this intuitive notion of iterative improvement with an iterative scheme. %
Its popularity is due  both to its simplicity and its stability as captured by the \emph{ascent} property \citep{dempster1977maximum, wu1983convergence}, which guarantees the iterative sequence $\{\theta^{(t)}, t=0, 1, \ldots \}$ (see the sidebar) will never lower the likelihood being maximized,  that is,
$\ell(\theta^{(t+1)}\mid X)\ge \ell(\theta^{(t)}\mid X)$, for all $t\ge 0$. %
The simplicity here is both computational and conceptual, because statisticians' familiarity with many standard complete data models make it easier for us to identify effective data augmentation schemes.

The idea of easier or more feasible optimization via introducing hypothetical data translates to Monte Carlo sampling problems.  The data augmentation (DA) algorithm of \cite{tanner1987calculation} epitomizes this idea, which has stimulated many of the subsequent refinements and generalizations  \cite[e.g.,][]{liu1999parameter,van2001art,pal2015improving}; see \cite{hobert2011data} for a review.  With a DA algorithm, sampling from a posterior density $\pi(\theta \mid X) \propto p(\theta)p(X \mid\theta)$ is augmented to sampling from $\pi(\theta \mid X, Z) \propto p(\theta)p(X, Z \mid\theta)$. One then implements the Gibbs sampler described in Section~\ref{sec:MC} by iterating between sampling from
$\pi(\theta \mid X, Z)$ and  from $p(Z\mid X, \theta)$, which are the sampling counterparts of the M-step and E-step, respectively.  %
 DA works whenever the introduction of $Z$ allows the construction of a Markov chain with stationary density $\pi(\theta,Z\mid X)$ that is considerably easier to run than the one with stationary density $\pi(\theta \mid X)$.   
Not surprisingly, there is a large class of statistical models for which both EM and DA are effective, for likelihood and Bayesian computation, respectively. There are also many parallels between EM-type of algorithms \citep[e.g., the ECM algorithm,][]{meng1993maximum} and MCMC algorithms in terms of both theoretical properties and implementation strategies, as detailed in \cite{van2010cross}.

\section{The Magic Touch: Refinement by Confinement}\label{sec:touch}

This sense  brings to the fore the advantage of creating ingenious model constraints and links, be they conceptual  or functional,  between different parameters. Bayesian pooling and shrinkage tether the parameters in the model and allow the transfer of information between groups of observations. %
This results in more information being available for each parameter estimate.   Similarly, the general concept of conditioning shrinks the sample space to relevant subspaces, and allows  the insertion of subject-matter knowledge into the mathematical aspects of the statistical analysis.

\subsection{Shrinkage Estimation and Bayesian Pooling}\label{sec:pooling}

Bayesian ideas have a long history that has been interweaved with frequentist and fiducial insights \citep[e.g.,][]{fienberg2006did}. Their importance is rooted in a series of attractive features that are available almost automatically and are backed by principles with firm theoretical support. In their modern form, Bayesian analysis and inference are fundamental to statistical modeling and to practical data analysis. 

One powerful expression of Bayesian modeling is the flexible \emph{pooling} of information.
In simplest terms, pooling is the act of borrowing strengths from individual constituents, to achieve something superior than what's possible by each of them. 
The James-Stein estimator \citep{stein1956inadmissibility,james1961estimation}, and in general shrinkage estimators, emphasize the importance of pooling in the simultaneous estimation of multiple means. Suppose that there are $M$ groups, each with population mean $\{d_j:j=1,\ldots,M\}$. One measurement is taken per group with standard Normal error: $Y_j \sim N(d_j, 1)$.
The MLE in this case is just $\hat{d}^{MLE}_j = Y_{j}$. As much as we expect it to hold many good properties,
it turns out that  whenever $M \ge 3$, the MLE is \emph{inadmissible}. The latter term signifies that the MLE can be easily dominated in terms of the \emph{risk} incurred in the estimation: 
$R_{L}\left({\bf d},\hat{{\bf d}}\left({\bf Y}\right)\right)=\mathbb{E}\left(L\left({\bf d},\hat{{\bf d}}\left({\bf Y}\right)\right)\right)$,
where expectation is taken with respect to the sampling distribution of the data ${\bf Y}$. 
The loss function $L$ can be measured in terms of the usual average squared error
\citep{james1961estimation}, but also under a more general class of convex functions \citep{brown1966admissibility}. 
The James-Stein estimator
\begin{equation}\label{eq:james-stein}
    \hat{d}^{JS}_{j} =\left(1-\frac{M-2}{\sum_{j=1}^{M} Y_{j}^{2}}\right)Y_{j},
\end{equation}
 improves the estimation risk everywhere and, in particular, has a lower risk than the MLE. 

\begin{marginnote}
\entry{Admissibility}{A statistical procedure is admissible if it is not completely dominated by another with respect to a given criterion.}
\end{marginnote}

The surprising result that MLE is not admissible can be explained rather intuitively from a data-augmentation perspective,  by considering a regression setting with $\{(Y_i, d_i): i=1, \ldots, M\}$ as augmented data (since $d_i$ is not observed). From this perspective, \cite{stigler19901988} argued that the MLE corresponds to regressing $Y$ on $d$ because $\mathbb{E}(Y_i\mid d_i)=d_i$. But this is the wrong regression, since what we really want is to predict $d_i$ from $Y_i$. Hence, we should regress $d_i$ on $Y_i$, which is in the form of $\beta Y_i$. Since $d_i$ is unobserved, the multiplier in front of $Y_i$ in \eqref{eq:james-stein} is an unbiased estimator of $\beta$ based on the $Y_i$'s alone \citep{stigler19901988,meng2005comment}.  

The James-Stein estimator itself is still not an admissible estimator for the population means, and is not proper Bayes. However, it can be motivated as an \emph{empirical Bayes} solution to the estimation problem, and serves as a basis for estimators with improved risk. \cite{strawderman1971proper} derived the class of minimax and admissible estimators, assuming the dimension $M \ge 5$, as the proper posterior mean under a class of Normal scale mixture priors.  The powerful result of \cite{brown1971admissible} established that proper and generalized Bayesian solutions constitute a complete class for the estimation of multivariate normal means under quadratic loss, further strengthening the connection between minimax shrinkage estimation and Bayesian procedures.

\begin{marginnote}
\entry{Empirical Bayes}{The practice of Bayesian inference where aspects of the prior distribution are determined by the data being analyzed.}
\end{marginnote}

Shrinkage estimators achieve uniformly superior estimation risk via an act of \emph{pooling} the observations towards a point or a subspace. The hierarchical Bayes models, also known as multilevel models, allows for the flexible pooling of information about subpopulation-specific parameters. Through the specification of the hyperpriors, the modeler controls for the extent of pooling performed on the constituents.  
Consider a simplified version of the tire example from Section~\ref{sec:info}, where we assume there is no brand effect,
and hence $b_i=0, i=1, \ldots, K$. We also absorb the baseline term
$a$ into the driver effects $d_j, j=1, \ldots, K$, so (\eqref{lm}) becomes $Y_{ij}=d_j+\epsilon_{ij}$ with $\epsilon_{ij} \overset{i.i.d.}{\sim} N(0, \sigma^2)$, and accordingly 
the prior from (\eqref{lmm}) becomes  $d_1,\ldots,d_M \overset{i.i.d.}{\sim} N(\mu_d, \tau_d^2)$.  
Here the value $\mu_d$ can be viewed as our prior knowledge of the average driver effects, and $\tau_d^2$ expresses the similarity between these effects.  %

Under this setup,  after observing  data $\vec{Y}_j=\{Y_{1j},\ldots, Y_{Kj}\}$, the posterior distribution for $d_j$ is 
\beq
d_j \mid \vec{Y}_j \sim N((1-\lambda_j) \bar Y_{\cdot j} + \lambda_j \mu_d,\ \tilde{\tau}_d^2), \quad j = 1,\ldots, M,
\label{pool}
\eeq
where $\bar Y_{\cdot, j}$ is the average of $\{Y_{1j},\ldots, Y_{Kj}\}$, and $\lambda_j$ is the \textit{pooling weight}, given by $\lambda_j = 
\tilde\tau_d^2/\tau_d^2$, which is the ratio of \textit{prior precision} to \textit{posterior precision}, where precision is defined as the reciprocal of the variance. Here the \textit{posterior precision} $\tilde\tau_d^{-2}$ itself has a very intuitive expression 
\beq
\underbrace{\tilde\tau_d^{-2}}_{\textbf{Posterior Precision}}=\underbrace{K\sigma^{-2}}_{\textbf{Data Precision}}+\underbrace{\tau_d^{-2}}_{\textbf{Prior Precision}}.
\label{postprec}
\eeq
Also, the term $K\sigma^{-2}$ is the data precision because it is the reciprocal of the variance of $\bar Y_{\cdot, j}$, our estimate of $d_j$ from data alone. 

The pooling attribute of the multilevel model (\eqref{pool}) is apparent, since the posterior mean of $d_j$ is a linear combination between the \emph{no pooling} mean (corresponding to $\lambda_j=0$, or $\tau_d^2 =\infty$) and the  \emph{complete pooling} one (corresponding to $\lambda_j=1$, or $\tau_d^2 = 0$).  In the former case, the prior average $\mu_d$  is useless for estimating $d_j$ because, by setting $\tau_d^2 =\infty$, we declare that there is no similarity whatsoever among the drivers' effects. Hence there is no information to borrow from other drivers (via the mean $\mu_d$), as captured by the prior precision of $\mu_d$ for estimating $d_j$ being zero.  Consequently, all our information about $d_j$ come from the data produced by the $j$th driver. In the latter case, by setting  $\tau_d=0$, we impose the assumption that all driver effects are the same, and since we have already specified that their average is $\mu_d$, each of them must be $\mu_d$, and hence no data would be needed or can help. In general, the value of $\lambda_j$ determines how much pooling towards the common $\mu_d$,  depending on how similar we consider the driver effects are: the more similarity, the smaller is $\tau^2_d$, and hence the larger $\lambda_j$, resulting in more polling.

In general, the posterior variance is smaller than the variance of the sample mean since it is ``shrunk'' by our prior knowledge/assumption of the similarity among the individual effect. The ``borrowing of information'' phenomenon described in the simple model above is a cornerstone of hierarchical Bayes models
\citep{gelman2013bayesian} that is useful in moderating the effect of extreme observations.  One of its most lauded features is  the improvement it brings to  predictions of cluster-level effects \citep[e.g.,][]{gelman2006multilevel} which, in our example,  are the driver-specific effects $d_j$, for all $1\le j\le K$.

\subsection{Conditioning}

As one of our favorite teachers of statistics likes to say, \emph{conditioning is the soul of statistics} \cite[p. 42]{blitzstein2015introduction}. Conditioning is an important operation that permeates methods under all schools of statistical inference: frequentist, likelihood, Bayesian, fiducial.

To condition on a random variable is to reduce the realm of possible outcomes to only those that yield the same value of the conditioning variable. By imposing such a confinement, the variability is on average reduced. Broadly speaking, we perform conditioning in order to sharpen the focus and provide an inferential conclusion with better \emph{relevance} to the question at hand \citep{liu2016there}. By way of analogy, when asked a question (especially a trick question), the sassy response would always open with ``it depends''. By asserting that the answer to the question is a function of the scenario under which we envision the answer, we are effectively performing conditioning. But precisely what kind of things should the answer depend on?

Bayesian inference is always conditioned on the entirety of the observed data. Since the data contributes to the Bayesian posterior  through the likelihood, the same thing can be said about likelihood inference. Often is the case that there exists a sufficient reduction to the data, that is, a function $t(X)$, such that the likelihood function depends on the data $X$ only through the sufficient statistic $t(X)$.

\begin{textbox}[t]\section{Stratification and Blocking}
In survey sampling and observational studies, \emph{stratification} refers to the method of obtaining samples from explicitly specified partitions of the intended population \citep{kish1965survey}. These partitions are usually determined by important features  of the population, such as people of the same age, sex, race and ethnicity, or other aspects that matter in the context of the study.  We employ stratification to make sure that each defined subpopulation is  represented in the sample precisely according to a defined proportion (usually the population proportion), without deviation due to randomness.  A similar idea, called \emph{blocking}, is encountered in experimental design \citep{box1978statistics}. Blocking is the deliberate effort on the experimenter's part to limit the random assignment of treatments in such a way that, again, each treatment arm is comprised of the defined subpopulations precisely according to a defined proportion. Both stratification and blocking are applications of the idea of \emph{conditioning}.
\end{textbox}

\begin{marginnote}
\entry{Ancillarity}{In likelihood theory, a statistic is ancillary to a parameter if it is a part of the parameter's minimum sufficient statistics but its distribution is free of the parameter.}
\end{marginnote}

Operating outside of the likelihood and Bayesian inference paradigms, however, the importance of conditioning on the sufficient statistics is most elegantly captured by the Rao-Blackwell theorem \citep{rao1945information,blackwell1947conditional}. The theorem establishes that, if $y(X)$ is an estimator for the parameter of interest $\theta$, the conditional expectation of $y(X)$ given a sufficient statistic of the parameter, say $t(X)$, will improve under any convex loss function. In other words, through conditioning, the sufficient statistic is capable of turning any estimator into a better one. The improved estimator, $\mathbb{E}\left(y(X)\mid t(X) \right)$, is unbiased if and only if the original estimator $y(X)$ is unbiased. If the sufficient statistic $t(X)$ is furthermore \emph{complete}, to Rao-Blackwellize the unbiased estimator results in a uniformly minimal variance unbiased estimator (UMVUE; \cite{lehmann1950completeness,lehmann1955completeness}), implying that it cannot be further improved upon under squared loss. The Rao-Blackwell theorem has important implications in the design of MCMC methods \citep{robert2021rao,kong2007further}, among many other applications. It provides a guideline and inspires methods to construct estimators with reduced Monte Carlo variability \citep[e.g.][]{liu1995covariance}.

In frequentist hypothesis testing,  conditioning on \emph{ancillary} statistics helps achieve another goal, that is to derive relevant conclusions based on the data at hand. Specifically, we wish to ensure that the conclusion enjoys the same quality -- as measured by test's statistical power  -- conditional on aspects that in themselves do not provide discriminative evidence towards the hypothesis. The idea is captured by the \emph{principle of conditionality} \citep{cox1974theoretical,reid1995roles}, which requires that the inference be drawn conditional on an ancillary statistic when one exists.  Consider the classic example given in  \cite{cox1958some}, recapitulated in \cite{fraser2004ancillaries}. A noisy measurement about an unknown parameter $\theta$ was taken under one of the two equally possible scenarios. Under the first scenario, $X \sim N(\theta, \sigma^2)$ and under the second scenario, $X \sim N(\theta, (100\sigma)^2)$, thus much noisier than the first. Notably, the indicator for which scenario was taken under is ancillary to the parameter of interest. For testing the null hypothesis of $\theta = 0$, the most powerful $95\%$ hypothesis test would call for a rejection rule that is  $|x| > 5\sigma$  if the experiment were conducted under the first scenario, and  $|x| > 164\sigma$ if under the second scenario. While this rejection region provides the greatest \emph{unconditional} statistical power for all $\theta' \neq \theta$, the merits of the conclusions under the two scenarios are vastly different. In particular, it has excellent power under the first scenario but is only mediocre under the second. In contrast, the \emph{conditional} test that would achieve the same statistical power under both scenarios would reject the null hypothesis when  $|x| > 1.96\sigma$ if the measurement were taken under the first scenario, and $|x| > 196\sigma$ if under the second. Even though the conditional test does not achieve the maximal unconditional power, the analyst would have the comfort of knowing that, regardless of which scenario they're working under, the quality of the resulting statistical conclusion is the same.

Conditioning can also be used to simplify more complex testing problems, particularly by getting rid of  nuisance parameters. Because the benefit of dimensionality reduction is too great,  it is sometimes performed at the expense of discriminative information about the parameter of interest. In the testing of independence from two-by-two contingency tables, Fisher's exact test is constructed conditional on the row and column margins. The row and column margins are not ancillary to the parameter of interest, here the cross-product ratio of the population proportion parameters of the four cells. Conditioning on them is justified on the grounds that table marginals contain very little information about the cross-product ratio \citep{yates1984tests}, and because it turns an otherwise three-dimensional problem into a unidimensional one; see \cite{little1989testing} for a more detailed discussion. Conditioning as a guiding principle is reflected  in practical techniques, such as \emph{stratification} in survey sampling and  \emph{blocking} in experimental design. The accompanying sidebar reviews these concepts.

\section{Just a Taste: Richness in Frugality}

Reading the title of the section one's thought goes towards the principle of parsimony, widely embraced and assiduously practiced by statisticians. However, the sense of frugality goes beyond that. As hinted in the introduction, the statistician is constrained to frugality by the very nature of the discipline and its methodology. Bootstrap demonstrates brilliantly that  within the  sample lies a wealth of information. The propensity score exemplifies frugality in the form of a single matching score that balances many observed attributes, and empowers the study's conclusions to go beyond the mere detection of association.  

\subsection{Bootstrap}

\emph{Bootstrap} \citep{eforn1979bootstrap} is a great illustration of the essence of randomized replications, as well as the caution needed to implement them correctly.  A set of observations $D_n=\{X_1,\ldots,  X_n \}$  does not automatically imply replications of any kind without further description of how they arrived at our desk or disk, regardless how large $n$ is or how many different values the observations may take.  For example, they could be just a single deterministic sequence out of a mathematical textbook.  But when we know, or more likely we assume, that it comes from a simple random sample (SRS) of size $n$, we can obtain  many (hidden) randomized replications.  For example, there are $n$ SRSs of size $n-1$,  denoted by $D_{n, -i}$, which are obtained by removing $X_i$ from $D_n$.   They are highly related to each other, but nevertheless, they are authentic randomized replications of each other because they are created by the same sampling process.

Now suppose we want to evaluate the variability of a statistic $g(D_n)$, such as the sample mean, $\sum X_i /n$.  If we could have many replications, say $D_n^1, \ldots, D_n^m$, we could estimate the variability by using the variability among $g(D_n^1),\ldots, g(D_n^m)$.  In practice, we only observe one $D_n$.  Nevertheless, if we permit ourselves to approximate $g(D_n)$ by $g(D_{n-1})$, then we  have at our disposal $n$ replications of $g(D_{n-1})$ via $g(D_{n, -i})$, $i=1,\ldots, n$,  with which we can assess the variability and other properties (e.g., bias) of $g(D_{n-1})$.  Historically, this is called the method of \emph{jackknife} \citep{quenouille1956notes,miller1964trustworthy}, and {bootstrap} is an improvement by avoiding the need to approximate $g(D_n)$ by $g(D_{n-1})$. By resampling from 
$\{X_1,\ldots,  X_n\}$ $n$ times with replacement,  bootstrap creates $n$ (approximate) randomized replications of $D_n$.

The intention of the bootstrap strategy is the same as the jackknife:  using the internal replicability of $D_n$ to create approximate randomized replications for itself.  In that regard, the bootstrap accomplishes an seemingly impossible task,  ``pulling ourselves up by our bootstraps'' by concocting apparently new data from of the old. On the surface, therefore, bootstrap might look like an act of ``double dipping'': from one sample, we create a large collection of resamples that allows one to build inference about yet a larger entity (the population). However, the nuanced theoretical justification of the bootstrap would make clear that what's being leveraged are the hidden (assumed) replications within the observed $D_n$, information that is
not fully utilized by the task of estimating $g(D_n)$ itself. The same idea has been put to good use in deriving other statistical methods, such as high-dimensional linear models \citep{zhao2021defense}.

But there is no free lunch.  Bootstrap is possible only if there indeed are hidden randomized replications in the sample.  Suppose the original $D=\{X_1,\ldots, X_n\}$ actually comes from a time series, that is, the index represents a  time order.  In such cases, removing one observation or sampling with replacement would destroy this dependence structure, and hence there is little reason to expect that the resulting bootstrap sample, if na\"ively constructed as before, would provide meaningful results.  However, if we know that the time series is stationary, that is, any continuous segment of the same size shares the same probabilistic behavior regardless of their locations, then effectively we have hidden randomized replications (in units of segments) to build upon. This leads to the idea of using \textit{block bootstrap} to sample (continuous) segments (i.e., blocks) of time series.
The theoretical justification of block bootstrap is more involved -- see \cite{buhlmann2002bootstraps} for an excellent overview of block bootstrap and other bootstrap methods for time series. But the fundamental idea is the same: using internal replicability to approximate  randomized replication.

\subsection{Propensity Matching}

Central to the analysis of randomized experiments and observational studies is the assignment mechanism, that is, the procedure through which units are allocated to different treatment groups. The key is to ensure that such an assignment will not introduce imbalances between the groups, which would make it difficult, if not impossible, to attribute differences in outcome to the treatments only.  When the assignments are not made probabilistically, as in virtually all observational studies, we make attempts to re-balance the two groups. For example, we may choose a subset of those who  received an existing treatment by matching their pre-treatment attributes to that of those who  received an experimental treatment according to some matching criteria.  Propensity matching \citep{rosenbaum1983central} is perhaps the most popular re-balancing method, partly because it offers a surprisingly practical and effective method for a seemingly impossible task, that is, maintaining balance statistically by matching only on a univariate score, the propensity score. 

The \emph{propensity score} of a unit $i$ is its probability to appear in the treatment group, where every unit is identified by its known covariates $X_i$ as well as its pair of potential outcomes $\{Y_i(1), Y_i(0)\}$, as introduced in Section~\ref{subsec:potential-outcome}. When the assignment mechanism, $Z_i \in\{0,1\}$, is (conditionally) \emph{unconfounded} \citep{dawid1979conditional}, 
namely the assignment received by unit $i$ is conditionally independent of its potential outcomes given the covariates, the propensity score  can simply be written as \citep{rosenbaum1983central,imbens2015causal}
\begin{equation*}
e\left(x\right)=\pr(Z_{i}=1\mid X_{i}=x, Y_i(1), Y_i(0)) = \pr(Z_{i}=1\mid X_{i}=x).
\end{equation*}
The propensity score plays a crucial role in the practical extension of causal analysis to observational studies. %
In classical randomized experiments, similarity among units (in the form of partial exchangeability) between the treatment arms can be established, more or less objectively, using assignment mechanisms under the control of the experimenter. Such luxury does not exist in observational studies, so to be able to conduct causal analysis, one must be able to decide whether two units are deemed similar based on available information. Thus, the criterion must be computable from the unit's available information, that is, their covariate information. This criterion that we need is the \emph{balancing score}, a function $b$ of the covariates $X_i$ with the property that if we condition on it,  all differences, if any, in the covariates no longer bias the assignment in any way: $Z_{i}\perp X_{i}\mid b\left(X_{i}\right).$
Trivially, the totality of the covariate itself is a balancing score:  $b\left(X_i\right) = X_i$. But matching on $X_i$, whereas desirable, is not practical especially when $X_i$ is of high dimension, because we would quickly run out of matching sample (e.g., to form a control group) since each person is unique.

The propensity score is the \emph{coarsest} balancing score, in the sense that it is a function of all other balancing scores. This analogy is precisely that, in the effort to de-bias assignment probabilities of units, the balancing score is a sufficient statistic of the unit's characteristics, whereas the propensity score is a univariate minimal sufficient statistic. That is, we  need only to match on a univariate quantity $e(X)$ in order to balance on the entire $X$ for estimating the treatment effect, a rather remarkable achievement.  
The propensity score allows for the estimation of the causal effect in observational studies through identifying \emph{comparable} units \citep{rosenbaum1984reducing,rubin1992characterizing}, when we do not have the luxury of controlled randomized trials which enforce such comparability. \\

\section{Sense and Sensibility}\label{sec:six}

We intend for this sense to be a culmination of our categories, in that it relies on all senses but allows the total ability to be larger than the sum.
Anything that we wish to predict or infer 
remains unknowable to us, unless we build a bridge between that and what we already know.
Yet, such bridges can never be fully tested before we embark on them, precisely because we rely on them to find out where they would lead us to. This ``bridge-building'' is the paramount challenge behind the practice of inference and prediction, and some ``leap of faith" is necessary for such statistical endeavors. Of course, there are different bridges one can build, and which one is better for which journey requires the ultimate statistical sense and sensibility.

When we are asked to solve a deterministic problem, say, to find the value of $\theta$ when we know 
\begin{equation}\label{eq:sense}
    X=\theta+U,
\end{equation}
there is only one bridge, that is, we solve it by first expressing $\theta$ as 
\begin{equation}\label{eq:solu}
    \theta=X-U,
\end{equation}
and then use whatever information we have about $X$ and $U$ to determine $\theta$. If $X=6$ and $U=0.1$, then $\theta$ must be $5.9$.  If $X=6$, but we don't know $U$ other than $|U|\le 0.1$, then $\theta$ will be restricted to an interval $[5.9, 6.1]$.  In either case, the link 
(\eqref{eq:sense}) permits us to transfer our \textit{deterministic}  knowledge, expressed as an equality or inequality, about $X$ and $U$ into either full or partial knowledge about $\theta$. 

Suppose now that our knowledge about $U$ is \textit{stochastic}, mathematically described by a probabilistic distribution: $U$ is uniformly distributed on $[-0.1, 0.1]$.  How can we transfer such stochastic knowledge to that about $\theta$?  This is where all the fun and frustration take place. We perhaps all have some vague sense that this distributional knowledge must impose some restrictions on $\theta$. The question then is how should we proceed \textit{sensibly}? 
At  first sight, this seems to be a rather trivial problem, at least for this toy example.  
Since $X=6$, and $U$ is uniform on $[-0.1, 0.1]$, should then $\theta $  be uniform on $[X-0.1, X+0.1]=[5.9, 6.1]$ as  (\eqref{eq:solu})  suggests?

\paragraph{Fiducial inference}

Above is the answer provided by the so-called \emph{fiducial inference}, put forward by R. A. Fisher \citep{fisher1935fiducial}, the founder of likelihood theory and many other statistical methods that are in use today. Fiducial inference is generally regarded as Fisher's biggest  blunder because the seemingly natural operation of mimicking the deterministic ``equation-solving'' step turned out to be rather problematic with stochastic quantities   \citep[see][]{zabell1992ra,dawid2022fiducial}. Even so, it influenced generations of statisticians and inspired fascinating topics such as structural inference \citep{fraser1968structural}, functional inference \citep{dawid1982functional}, Dempster-Shafer theory \citep{dempster1966new, shafer1976mathematical}, generalized fiducial inference \citep{hannig2016generalized}, and inferential models \citep{martin2015inferential}.  A fundamental issue is that, whereas the distributional assumption on $U$ induces a distribution for $X$ for any given value of $\theta$ in (\eqref{eq:sense}), it says nothing about whether $\theta$ can be described by a distribution or not.  If $\theta$ is not even endowed with a distributional structure, then what does it mean to say $\theta$ is \textit{distributed} uniformly on $[X-0.1, X+0.1]$? If it is given a distributional description, e.g., the so-called prior distribution $\pi$, then how $\theta$ is distributed after knowing $X$ will depend on how we specify $\pi$.  For example, if $\pi$ is uniform on $[6, 7]$, then it is impossible for $\theta$ to be uniformly distributed on $[5.9, 6.1]$ after observing $X=6$ because any value below 6 has already been eliminated by the prior $\pi$.

\begin{marginnote}
\entry{Fiducial (\emph{adj.})}{assumed as a fixed basis for comparison. Comes from the Latin word which means  ``trust''. (Dictionary.com)}
\end{marginnote}

The fiducial argument relies on an explicit leap of faith, in that it \emph{continues to regard} \citep{dempster1963direct} the distribution of the auxiliary variable as unchanged after the observation.  In the toy example,  this means to steadfastly treat $U$ as uniform on $[-0.1, 0.1]$ even after observing $X=6$. The argument seems to be sensible intuitively.  If we have not been given any prior knowledge or constraints about $\theta$, then whatever information generated by $X$ via the unknown prior $\pi$ about $U$ is not quantifiable, and hence we can ignore it, and persistently use the pre-data distribution information about $U$.  Unfortunately, this argument cannot be coherently rationalized and operationalized in terms of probabilistic distributional calculations.
This is because there is no probability distribution to describe \emph{ignorance}: any probability distribution specification about $\theta$ encodes restrictions on how likely  one possible state of $\theta$ is versus another, and hence it cannot represent ignorance; see below.  The situation is a bit like doing arithmetic without the number zero, and uses some other (small) numbers to represent zero, which may be acceptable in some practical cases but logical inconsistencies and paradoxes would  inevitably ensue (e.g., only a true zero remains zero after multiplication by any number). %

\begin{figure}
\begin{multicols}{3}
\centering{\includegraphics[width=0.3\textwidth]{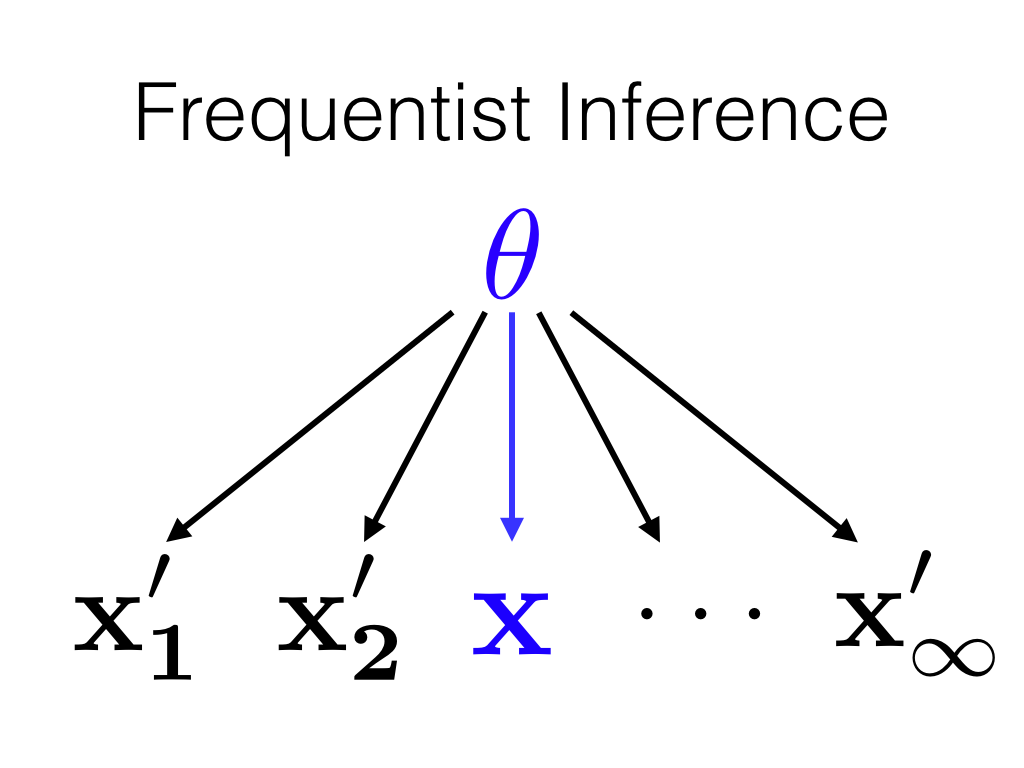}} 
$$ p(\mathbf{X}'|\theta)$$ \\ \phantom{text} 
\centering{\includegraphics[width=0.3\textwidth]{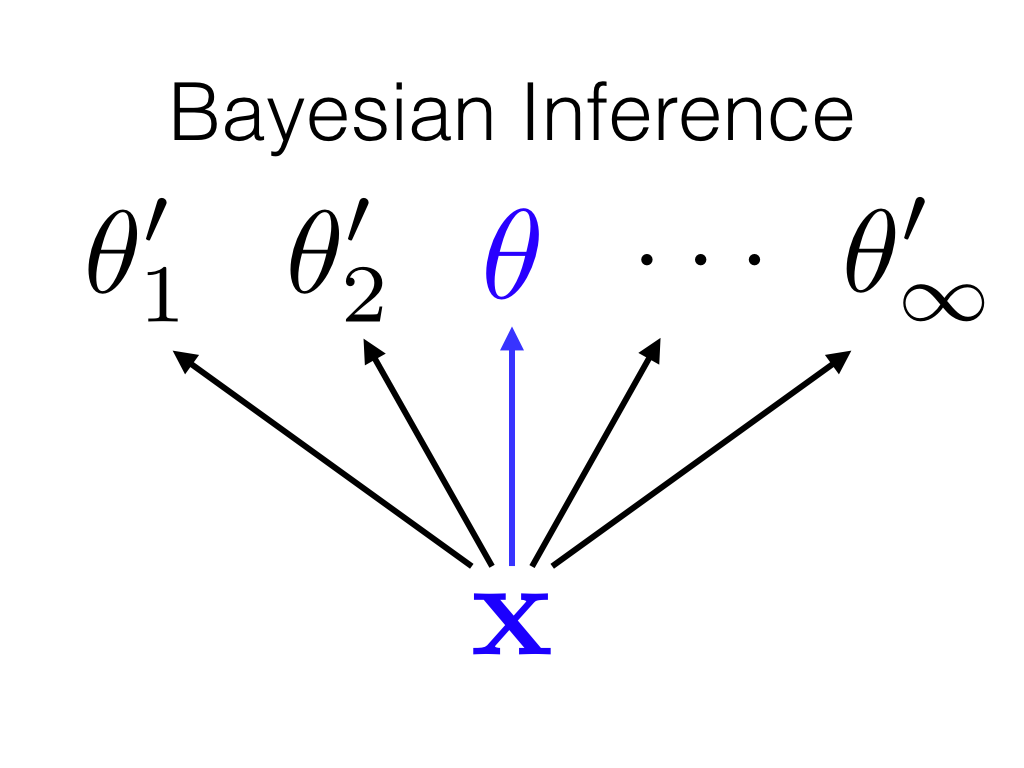}}
$$ p(\theta'|\mathbf{X})$$ \phantom{text} 
\centering{\includegraphics[width=0.3\textwidth]{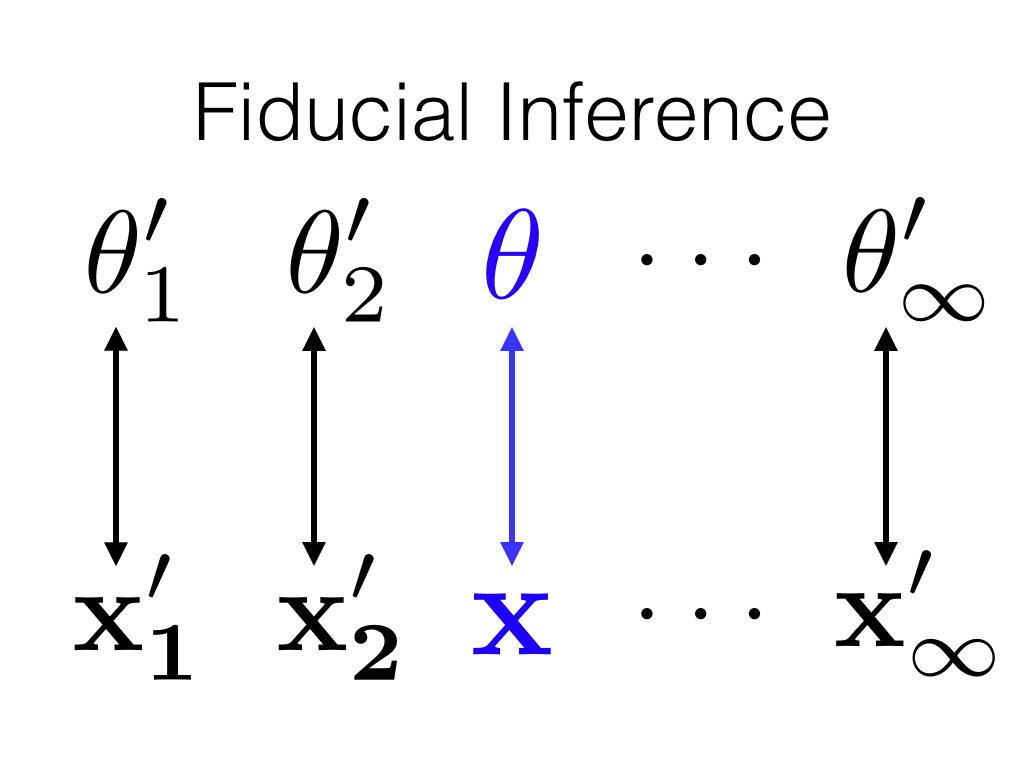}}
$$ p(\mathbf{X}',\theta'|U'(\mathbf{X'}, \theta')=U(\mathbf{X},\theta))$$ \\ \phantom{text} 
\end{multicols}
\caption{Replications underlying the frequentist, Bayesian, and Fiducial inferential paradigms.  $X$ is the data,  $\theta$ is the estimand, and $p$ is the respective conditional distribution.} 
    \label{fig:replication}
\end{figure}

\paragraph{Bayesian inference}
The Bayesian paradigm invokes a different kind of leap of faith, which is to require a prior distribution for $\theta$.  In reality, we may have very good knowledge to impose, say, $5 < \theta < 6$, but almost never have full confidence to say how $\theta$ is distributed over the interval $(5, 6)$. This is even the case for those of us who are comfortable with the idea of using a probability distribution as a mathematical tool to describe our uncertainties (and hence avoid the issue of considering $\theta$ to be random). It is extremely rare, if not impossible, to know precisely the relative uncertainty about one possible state of $\theta$ versus another, and for all such pairs. Nevertheless, once a prior for $\theta$ is posed,  it allows us to create the most relevant replications $(\theta', X)$ for drawing inference about $\theta$ that corresponds to our actual data $X$,
as depicted in the middle panel of Figure~\ref{fig:replication}. It is the most relevant because  replications $(\theta', X)$ all share the same data $X$ as observed, and hence the posterior distribution $p(\theta'\mid X)$ directly addresses our inference question: how likely is each possible value $\theta$, denoted by $\theta'$, given our data $X$?

The concern of the unreliability of arbitrary priors has led to the quest for formal rules to guide their choice \citep{kass1996selection}. The quest gave rise to the literature of \emph{objective priors} \citep[e.g.][]{berger2008objective,ghosh2011objective,berger2012objective,berger2015overall}, or non-informative priors as generally known. A specific class of objective priors is the so-called \textit{reference priors}, which encompasses priors that maximize some chosen distance between the prior and posterior distributions \citep{berger1989estimating,ye1993reference,berger2009formal}. The rationale of this maximization is to ensure that the likelihood function has the most impact, or to let the data speak as loudly as possible. Historically, the concern of not being able to choose a sensible prior was addressed by requiring a statistical procedure to be reliable regardless of the choice of the prior \citep{neyman1934two}, which is the same as requiring reliability when the prior is a point mass at any plausible value of $\theta$, as described next.

\begin{marginnote}
\entry{Objective Bayes}{A somewhat paradoxical term describing efforts to invoke a prior while minimizing its impact.}

\end{marginnote}

\paragraph{Frequentist inference}
The frequentist paradigm corresponds to a top-down
replication postulate, which  considers all possible data sets that share the same $\theta$ value as the actual observed dataset, as illustrated by the left panel in Figure~\ref{fig:replication}. The frequentist approach then investigates the long-run frequency properties of various methods over such hypothetical datasets, denoted by $X'$, and then chooses one with the desirable properties, such as being unbiased.  A key feature of this  approach is that it entirely bypasses the need to specify a distribution for $\theta$. To some, this represents scientifically a more ``objective'' paradigm for inference. However, the no-free-lunch principle tells us that there must be a catch. 
The long-run frequency properties are assessed by considering how our procedures perform on these hypothetical $X'$s, such as coverage probability of a confidence procedure, or power of a hypothesis testing procedure. Whether such properties can  guarantee anything about the procedures' reliability on the actual $X$ we observe is fundamentally a matter of transferring faith about their collective merits into individualized performance. To some, this is an even more dangerous leap of faith than either the Bayesian one or the fiducial one, since the latter can be viewed as creating replications on the joint space $(X', \theta')$, subject to the constraint that their ``noise'' $U'$ must be the same as the $U$ corresponding to the one underlying our actual data,
as depicted in the right panel of Figure~\ref{fig:replication}. 
\medskip

But whichever replications we prefer, there is no escaping the leap of faith. Putting it differently, ultimately it is our ``sixth sense''  that completes our journey from the known to the unknowns. The differences between paradigms are more a matter of what we consider sensible, a question whose universal answer is no easier than establishing  \emph{a theory of everything}.   \\

\section{How Did We Form Our Senses? }

Our ``sensical tour'' started with a pleasantly surprising invitation from the editors of this journal to write about ``The Top X Big Ideas in Statistics''. Given the long history and yet the accelerating evolution of statistics,  where does one even begin with such a task?

Being all trained quantitatively but with appreciation for qualitative thinking, we first pondered on the meaning of almost every word in the ask. In this context, \textit{Top X} seems to be a sure invitation for controversy because it demands  a univariate ordering of a high-dimensional enterprise.  \textit{Big} is also a big trouble:  big on what? Ingenuity, insight, influence, impact, or all of these?  And do we mean \textit{ideas} only, as products of human intelligence, not \textit{discovery},  as secrets of God (or nature) revealed by humans?  Two central pillars of statistics and probability are the \textit{law of large numbers} and the \textit{central limit theorem}. Neither is an idea. Whether or not  human beings find its mathematical expression, the bell curve would still reveal itself everywhere from Galton's boards to the score boards. 

To further complicate the matter,  \textit{ideas in statistics} are not the same as \textit{statistical ideas}, which need not be in statistics. A recent example is the use of \textit{data augmentation} for incorporating prior knowledge or model considerations by creating synthetic training samples that reflect them
\citep[see, e.g.,][]{shorten2019survey, taylor2018improving}. This is a fine statistical idea, a good example of using randomized replications to operationalize probabilistic modeling. But, because it is proposed in the machine learning literature, it is unclear whether there has been an awareness in that literature of the terminology clash with the statisticians' own data augmentation, reviewed in Section~\ref{sec:seeing}, let alone any discussion of the similarities and differences between the two usages of the same term \citep[e.g., no mention of the statistical usage in the review article by][]{shorten2019survey}.  

Ultimately we were inspired by the wisdom behind Stigler's ``Seven Pillars of Statistical Wisdom'' \citep{stigler2016seven} to accentuate the essence of statistics not by top or big ideas or discovery, but by its broad conceptual framing and its intellectual axes.
We choose to categorize our framing
into the six senses of statistics for reasons listed in Section~\ref{sec:senses}. Evidently, our own intellectual trajectories led us to the particular set of examples used to illustrate each sense. Space constraints have limited the scope of illustrations even further. We would have loved to include other topics that have colored our relationship with statistics. Indeed, we only touched upon a few of the seven pillars of \cite{stigler2016seven}:  (1) aggregation,  (2) information measurement, (3) likelihood, (4) intercomparison, (5)  regression, (6) experimental design, and (7) residual. The overlap between our example topics with those listed in ``What are the Most Important Statistical Ideas of the Past 50 Years?'' \citep{gelman2021most} is of a similar order: 
counterfactual causal inference, bootstrap and simulation-based inference, overparameterized models and regularization, Bayesian multilevel models, generic computation algorithms, adaptive decision analysis, robust inference, and exploratory data analysis. 

Such comparisons should make clear that, whereas we believe the six senses are essential for the statistical enterprise, there are many more important methods and ideas than those mentioned in this article (or any single article). The more mathematically inclined reader may have also noticed that we have steered clear of probability concepts (e.g., central limit theorems, ergodic theory, asymptotics) that are fundamental to statistical derivations and justifications. They are, in our opinion, the tools we need to advance once our senses are pointing us in the right direction, but they do not define a statistician’s \emph{phronesis}. Nevertheless, we hope that the examples used will facilitate the reader's ability to sharpen their statistical senses and guide discovery of new links with their favorite methods.

\begin{marginnote}
\entry{Phronesis (n.)}{“wisdom in determining ends and the means of attaining them, practical understanding, sound judgment" (Dictionary.com)}
\end{marginnote}

We are writing this at a time when the discipline of statistics is expanding rapidly under important stimulants such as the emergence of data science as an ecosystem \citep{Meng2019Data, meng2019five}.  Statistical ideas and principles are meshing with computational algorithms to solve scientific and societal problems that carry the burden of enormous, messy, and often confidential data, the responsibility of dealing with complex relationships, the increased darkness of the ``black boxes'' promoted by  advances in machine learning, and  the societal demands of transparency and interpretability due to fairness considerations, to name only a few. New senses therefore are likely to emerge, but those discussed in this article are time-honored, and will continue to play an important role in the way new problems are tackled. 

For example, the time-honored bias-variance trade-off manifests itself in the relevance-robustness trade-off \citep{liu2016there} in the context of accumulating statistical evidence for assessing the effectiveness of individualized treatments, an increasingly common desire due to the availability of the (seemingly) big data. Since no two individuals are identical, whether as  biological or social beings, evidence from proxy individuals are approximations in nature.  The more resemblance between the proxy individuals (or training sample)  and the target individual, the more relevant are the approximating assessments. But this relevance comes at the expense of fewer available proxies, resulting in  assessments that are less 
robust.  Conversely, we can assemble many proxy individuals if we relax on the proxy criterion and hence have more stable statistical assessment.  But the results may not be that relevant for the target individual because of the loose criterion employed in choosing the proxies.  

Whereas the optimal construction of the training sample is a Holy Grail in data science,  the kind of statistical phronesis that is discussed in this article 
helps us make a sensible treatment decision. As an extreme example, insisting on having  fully robust evidence is similar to insisting on  100\% coverage, which would then lead to a useless -- tautological -- confidence interval. But by giving up a small amount of robustness, we can achieve much better relevance,  just as we obtain a more meaningful confidence interval by giving up a small amount of confidence (e.g., 5\%). 

Those with good statistical senses are likely to possess more confidence in their abilities to handle whatever the future may bring. Our belief is built on the observation (or our sixth sense?) that statistical principles have demonstrated   enduring importance and influence in every ``scientific revolution'', be it big data mining, machine learning, data science, or the $n$th Spring of AI.

\bibliography{reference,copssbib,superref}
\bibliographystyle{ar-style1.bst}

\end{document}